\documentclass[mathematics,article,accept,pdftex,moreauthors]{Definitions/mdpi}

\newcommand{\REF}[1]{#1}
\newcommand{\Nu}{\mathrm{\it Nu}}

\firstpage{1} 
\pubvolume{1}
\issuenum{1}
\articlenumber{0}
\pubyear{2024}
\copyrightyear{2024}
\datereceived{ } 
\daterevised{ } 
\dateaccepted{ } 
\datepublished{ } 
\renewcommand{\vec}[1]{\boldsymbol{#1}}
\usepackage{natbib}

\hreflink{https://doi.org/} 

\Title{Modelling the transition from shear-driven turbulence to convective turbulence in a vertical heated pipe }

\TitleCitation{Title}

\Author{Shijun Chu $^{1}$, 
Elena Marensi $^{2}$ and
Ashley P. Willis $^{1,*}$
}

\AuthorNames{Shijun Chu, Elena Marensi and Ashley P.\ Willis}

\AuthorCitation{Chu, S.; Marensi, E.; Willis, A.P.}

\address{%
$^{1}$ \quad Applied Mathematics, School of Mathematical and Physical Sciences, University of Sheffield, Sheffield S3 7RH, UK.\\
$^{2}$ \quad School of Mechanical, Aerospace and Civil Engineering, University of Sheffield, Sheffield S1 3JD, UK.}

\corres{Correspondence: a.p.willis@sheffield.ac.uk}

\abstract{Heated pipe flow is widely used in thermal engineering applications, but the presence of buoyancy force can cause intermittency, or multiple flow states at the same parameter values.  Such changes in the flow lead to substantial changes in its heat transfer properties and thereby significant changes in the axial temperature gradient.
We therefore introduce a model that features 
a time-dependent background axial temperature gradient, and consider two temperature boundary conditions -- fixed temperature difference and fixed boundary heat flux.
Direct numerical simulations (DNS) are
based on the pseudo-spectral framework,
and good agreement is achieved between present numerical results and experimental results. 
The code extends openpipeflow.org 
\cite{willis2017openpipeflow} 
and is available at the website.
The effect of the axially periodic domain on flow dynamics and heat transfer is examined, using pipes of length $L=5D$ and $L=25D$.  
Provided that the flow is fully turbulent, results show close agreement for the mean flow and temperature profiles, and only slight differences in root-mean-square fluctuations.
When the flow shows spatial intermittency, heat transfer tends to be over estimated using a short pipe, as shear turbulence fills the domain.  This is particularly important when shear turbulence starts to be suppressed at intermediate buoyancy numbers.
Finally, at such intermediate buoyancy numbers we confirm that the decay of localised shear turbulence in the heated pipe flow follows a memoryless process, 
 similar to that in isothermal flow.  While isothermal flow then laminarises,  convective turbulence in the heated flow can intermittently trigger bursts of 
shear-like turbulence.
}


\begin{document}


\section{Introduction}
In the heated flow context, 
flow driven by an external pressure gradient is referred to as `forced' flow, while buoyancy resulting from the expansivity of the fluid close to a heated wall can provide a force that partially or fully drives the flow, referred to as `mixed' or `natural convection' respectively. 
In a model, buoyancy may only need to counter drag forces in the vertical pipe.  In practice, we are likely to encounter what could be called `super-natural' convection, where the buoyancy must be larger than the local drag in order to drive flow in a wider circuit.  In this case, flow in the vertical section of the circuit is subject to a reversed pressure gradient that limits the flow rate.

Turbulent mixed convection in a vertical pipe is a representative model for heat transfer that can be found in thermal engineering applications, e.g.\ heat exchangers, nuclear reactors, chemical plants and cooling systems for electronic components \citep{zhang2020review}. Despite the relatively simple geometry, the flow state and heat transfer can be difficult to predict in the presence of buoyancy. Buoyancy can enhance the heat transfer in a heated downward pipe flow, but suppress heat transfer in upward heated pipe flow \citep{mceligot1970relaminarization,ackerman1970pseudoboiling,bae2005direct,wibisono2015numerical,zhang2020review}. In an upward pipe flow, with the enhancement of heating,  heat transfer first deteriorates slowly, then suddenly drops when shear-driven turbulence collapses, then recovers, and finally can approach as large values as for downward flow at large buoyancy parameters \citep{zhang2020review}. 

Heat transfer presents some complicated features in upward heated pipe flow, as well as the flow dynamics. Previous research has confirmed three flow states in different heating conditions and Reynolds numbers, i.e.\ shear turbulence, the laminar state and convective turbulence \citep{marensi2021suppression,chu2024minimal}. The laminar state can persist up to Reynolds numbers of around 3000, versus approximately 2000 in isothermal flow.
The addition of buoyancy suppresses and can laminarise shear turbulence. Research on the phenomenon of laminarisation in mixed convection can be traced at least as far back as Hall $\textit{et al.}$ \cite{hall1969laminarization}, which provided a theoretical explanation of this phenomenon, suggesting that reduced shear stress in the buffer layer leads to a reduction or even elimination of turbulence production.  More recently, He $\textit{et al.}$ \cite{he2016laminarisation} modelled the buoyancy with a radially dependent axial body force added to isothermal flow, successfully reproducing the laminarisation phenomenon. They found that the body force makes little change to the key characteristics of turbulence, and proposed that laminarisation is caused by the reduction of the `apparent Reynolds number', which is calculated based only on the pressure force of the flow (i.e. excluding the contribution of the body force). Similar laminarisation phenomena have also been observed for the isothermal case
in the presence of a modified base flow
\citep{hof2010eliminating,kuhnen2018destabilizing}.  It is conjectured that a flattened velocity profile reduces transient growth \citep{scheele1962effect}, thus suppressing shear turbulence. Chu $\textit{et al.}$ \cite{chu2024laminarising} examined the self-sustaining process \citep{hamilton1995regeneration} 
in this context
and found that the flattened velocity profile can suppress the 
instability of streaks
thereby disrupting the self-sustaining process of shear turbulence. 

There is a developed history of numerical simulations of mixed convection in vertical pipe flow using various methods.
In an early study, a modification of the Redichardt eddy diffusivity model was used to simulate mixed convection \citep{hiroaki1973effects}, but it proved that this approach did not adequately account for certain
local features of the flow. Cotton $\textit{et al.}$ \cite{cotton1986theoretical} used the low-Reynolds number $k-\epsilon$ turbulence model of Lauder $\textit{et al.}$ \cite{launder1974application} to simulate the vertical heated pipe flow
with some success. Behzademhr $\textit{et al.}$ \cite{behzadmehr2003low} conducted a study of upward mixed convection in a longer pipe at two rather low Reynolds numbers ($Re=1000$ and $1500$) over a range of Grashof numbers, which measures the heat flux at the wall, using the Lauder-Sharma model. They identified two critical Grashof numbers for each Reynolds number, which correspond to laminar-turbulent transition and relaminarisation of the flow. 
More recently, direct numerical simulation (DNS) has been used in studies of mixed convection. Kasagi $\textit{et al.}$ \cite{kasagi1997direct} conducted a DNS study at $Re=4300$ and several values of Grashof Number. The simulations show that buoyancy changes the distribution of Reynolds shear stress and shear production rate of turbulent kinetic energy, leading to heat transfer enhancement or suppression. You $\textit{et al.}$ \cite{you2003direct}  also did the DNS for the mixed convection in vertical pipe flow, and compared the results of upward and downward flow. Kim $\textit{et al.}$ \cite{kim2008assessment} presents an assessment of the performance of a variety of turbulence models in simulating buoyancy-aided, turbulent mixed convection in vertical pipes. They found the use of different methodologies for modelling the direct production of turbulence through the direct action of buoyancy has been shown to have little effect on predictions of mixed convection in vertical flows. Chu $\textit{et al.}$ \cite{chu2016direct} applied a well-resolved DNS to investigate strongly heated airflow in a vertical pipe at $Re=4240 $ and $ 6020$. The results showed an excellent agreement in heat transfer and flow statistics.  
Recent calculations at larger flow rates include 
\cite{antoranz2015numerical,straub2019influence,cruz2024high}. 

We wish to examine the detailed transient nature of transition, for which accurate DNS is necessary,
and since the flow type ultimately affects the heat transfer and hence the heating of the fluid itself, we wish to explicitly include a time-dependent temperature gradient.
The model 
developed in 
Marensi $\textit{et al.}$ \cite{marensi2021suppression}
extended the pseudo-spectral code openpipeflow \citep{willis2017openpipeflow}
to include a time-dependent spatially uniform heat sink.  This form for the sink has the advantage of a simple analytic expression for the laminar state.  Numerical results showed good agreement with experimental results, but were improved slightly in Chu $\textit{et al.}$ \cite{chu2024minimal} by associating the heat sink with a time-dependent background temperature gradient along the axis of the pipe.  
In both
Marensi $\textit{et al.}$ \cite{marensi2021suppression} and 
Chu $\textit{et al.}$ \cite{chu2024minimal}, 
fixed temperature conditions were used 
at the wall.
In this work, we provide further details on the model of Chu $\textit{et al.}$ \cite{chu2024minimal} and add a second case for the temperature boundary condition, that of fixed heat flux at the wall. 

\REF{It should be noted that our model assumes axial periodicity, which implies that it should be applied to a straight section of pipe, downstream of effects from an inlet or bend.  This approximation is widely adopted for research in shear turbulence \citep{eggels1994fully,avila2010transient} and mixed convection \citep{cruz2024high} in pipe flows. Another potential limitation is the Boussinesq approximation \citep{gray1976validity,turner1979buoyancy} adopted in our model, which ignores the effect of heating on viscosity and assumes that changes of density only need be considered in the buoyancy force term in Navier-Stokes equations. Nevertheless, such modelling simplifies the simulation greatly and provides good results in many circumstances \citep{gray1976validity},
and has been widely adopted in the simulations of mixed convection \citep{su2000linear,you2003direct,cruz2024high}.  As we focus on flow and heating rates that are transitional with respect to flow regimes, we do not consider extreme parameter values here.   When the Boussinesq approximation holds, there is mathematical equivalence between upward heated and downward cooled flow, i.e.\ the case modelled here could be experimentally examined by considering a hot fluid flowing down a pipe through a cold room.  Although the temperature along the pipe will approach the room temperature exponentially under such circumstances, it can be modelled to be locally linear over a reasonable distance, and the temperature gradient along the pipe will depend on whether the flow is laminar or turbulent.  
Finally, it should also be noted that
turbulence increases friction drag and hence pumping costs.  The relative importance of this cost is very context-specific, and therefore is not considered here.  Our focus is on the enhanced heat transfer due to turbulence.
}

 The plan of the paper is as follows. In \S 2, we present the model for DNS of vertical heated pipe flow, including two types of temperature boundary conditions, i.e. fixed temperature difference and fixed boundary heat flux. In \S 3, we first show the results of DNS, then present the results of different lengths of pipe.  Next, we show how the lifetime of shear turbulence changes with buoyancy force. Finally, the paper concludes with a summary in \S 4.

\section{Model for heated pipe flow}

 Let $\vec{x}=(r,\phi,z)$ denote cylindrical coordinates within a pipe of radius $R$. The total temperature satisfies
 \begin{equation}
    \label{eq:tottem}
    \frac{\partial T_{tot} }{\partial t}+({\vec{u}_{tot}} \cdot \vec{\nabla}) T_{tot}=\kappa \vec{\nabla} ^{2}T_{tot} \, , 
 \end{equation}
 where $\kappa$ is the thermal diffusivity.
We decompose the total temperature as  
\begin{eqnarray}
   \label{eq:tempexpansion}
    T_{tot}(\vec{x},t)&=&T_w(z,t) + T(\vec{x},t) - T_0 \, , \\
   \label{eq:Tw}
    T_w(z,t) &=& a(t)\,z + b \, ,
\end{eqnarray}
where  $a(t)$ is the 
time-dependent axial temperature gradient, $b$ is a constant reference temperature, $T(\vec{x},t)$ carries the temperature fluctuations,
and $T_0$ is a constant that will be used as a temperature scale.  
The factor $-T_0$ has been inserted in  (\ref{eq:tempexpansion}) so that the temperature fluctuations $T$ are positive and largest at the hot wall.
The bulk temperature we write
\begin{equation}
   \label{eq:Tb}
    T_b   =  \langle  T \rangle \, ,   
\end{equation} 
where the angle-brackets denote the 
volume average.  The important quantity that measures the heat flux is the Nusselt number
        \begin{equation}
         \Nu
         =\frac{2R\, q_w}{\lambda \, ({T|_{r=R}-T_b})}\,.
         \label{eq:Nu}
        \end{equation}
where $\lambda$ is the thermal conductivity 
and 
$q_w=\lambda\,\overline{(\partial T/\partial r)|_{r=R}}$ is the heat flux at the wall,
where the overline denotes the time average.  

Note that $\Nu$ is an observed quantity, rather than a prescribed parameter, as it depends on the state of the flow.

For the fixed temperature boundary condition, $T_w$ is the value of the temperature at the wall. Evaluating (\ref{eq:tempexpansion}) at the wall gives $T|_{r=R}=T_0$.  
The wall temperature is locally isothermal (does not deviate from $T_w$), while the heat flux may exhibit variations.
However, $q_w$ can be measured, and is expected to  be statistically steady, except when interrupted by a change of state of the flow, such as from shear turbulence to convective turbulence.

For the fixed heat-flux boundary condition, $q_w$ takes the same value everywhere.  
Local variations in the boundary temperature are possible, 
so that here,
$T_w$ represents an averaged wall temperature.  Note that $\Nu$ will still vary through changes in $T_b$.

Throughout the rest of this work,  dimensionless variables and equations are presented, except in the definition of the scales and dimensionless parameters.  We use $R$ as the length scale and twice the bulk flow speed $2U_b$ for the velocity scale, which for isothermal laminar flow coincides with the centreline speed.  For the temperature scale we use $T_0$, which will be linked to the boundary conditions in the following sections.  Using these scales we arrive at the dimensionless governing equation 
\begin{equation}
 \label{eq:nondimenTem}
        	\frac{\partial T }{\partial t}+({\vec{u}_{tot}} \cdot \vec{\nabla}) T =\frac{1}{Re\,Pr}\vec{\nabla} ^{2}T -{\vec{u}_{tot}}\cdot{\hat{\vec{z}}}\,a(t) ,
\end{equation}
where it is assumed that variations in the temperature gradient are much slower than variations in the local fluctuations, i.e.\ 
$\partial_t a(t)\ll \partial_t T(\vec{x},t)$.
The dimensionless parameters are the Reynolds and Prandtl numbers $Re=2U_bR/\nu$ and $Pr=\nu/\kappa$, where $\nu$ and $\kappa$ are the kinematic viscosity and thermal diffusivity. 
A Prandtl number of 0.7 is used in all calculations.  
The last term on the right-hand side is a sink term that withdraws the energy that enters through the boundary.  The value for  $a(t)$ at each instant is determined via the spatial average of (\ref{eq:nondimenTem}) and 
depends on the boundary condition on the temperature, as shown in the following sections.
Axial periodicity over a dimensionless distance $L=2\pi/\alpha$ will be assumed
for the temperature fluctuation field $T(\vec{x},t)$.

Axial periodicity is also assumed for
\REF{the velocity field $\vec{u}_{tot}(\vec{x},t)$.} Under the Boussinesq approximation \citep{turner1979buoyancy}, the dimensionless Navier--Stokes (NS) equations are
        \begin{equation}
        	\label{eq:NSE-total}
        	\frac{\partial\vec{u}_{tot}}{\partial t}+(\vec{u}_{tot}\cdot\vec{\nabla}) {\vec{u}_{tot}}=-\vec{\nabla} p + \frac{1}{Re}\vec{\nabla}^{2} {\vec{u}_{tot}}+\frac{\gamma g R T_0}{(2U_b)^2} T\hat{\vec{z}} + \frac{4}{Re}(1+\beta(t))\hat{\vec{z}} \, ,
        \end{equation}
     with continuity equation
        \begin{equation}
       	\vec{\nabla} \cdot \vec{u}_{tot}=0 \, ,
       \end{equation}
    and no-slip condition 
    $\vec{u_{tot}}=\vec{0}$  
    at the wall,
            where $\gamma$ is the thermal expansivity and $g$ is acceleration due to gravity.
    Here, 
    $\langle \partial_z p \rangle=0$
    and the non-zero component of the axial pressure gradient appears in the final term of (\ref{eq:NSE-total});
    $\beta(t)$ is the excess pressure fraction, relative to isothermal laminar flow, required to maintain the fixed dimensionless mass flux $\langle 
    {\vec{u}_{tot}}\cdot{\hat{\vec{z}}}
    \rangle=1/2$.  
    Further decomposing the variables as
    \begin{eqnarray}
    {\vec{u}_{tot}}(\vec{x},t)&=&u_0(r)\hat{\vec{z}}+\vec{u}(\vec{x},t) \, , \quad u_0=1-r^2 \, ,  \\  
    \label{eq:Tdecomp2}
    T(\vec{x},t)&=&\Theta_0(r) +\Theta(\vec{x},t) \, , 
    \quad \Theta_0 = r^2 \, ,    
    \end{eqnarray}
leads to governing equations for 
\REF{the deviation fields}
$\Theta$ and $\vec{u}=(u_r,u_\phi,u_z)$
\begin{eqnarray}
           	\label{T-perturbation}
        	\frac{\partial \Theta }{\partial t}+u_0\frac{\partial \Theta }{\partial z}+u_r\frac{d\Theta_{0}} {dr} + (\vec{u}\cdot\vec{\nabla}) \Theta =\frac{1}{RePr}\vec{\nabla} ^{2}\Theta +\frac{4}{RePr}-(u_0+u_z)a(t) \, , 
            \\
        	\label{NSE-perturbation}
        	\frac{\partial \vec{u}}{\partial t}+u_0\frac{\partial \vec{u}}{\partial z}+u_{r}\,\frac{du_0}{dr}\, \hat{\vec{z}}+(\vec{u}\cdot \vec{\nabla}) \vec{u} =   -\vec{\nabla}  p + \frac{1}{Re}\vec{\nabla} ^{2}\vec{u}+\frac{4}{Re}(C (\Theta+\Theta_0)+\beta(t))\hat{\vec{z}} \, ,
\end{eqnarray}
with continuity condition $\nabla\cdot\vec{u}=0$ and boundary condition $\vec{u}=\vec{0}$.
The parameter $C$ measures the buoyancy force relative to the pressure gradient for laminar flow.  Equating buoyancy terms in (\ref{eq:NSE-total}) and (\ref{NSE-perturbation}), we have
\begin{equation}
   \label{eq:Cscales}
    \frac{4}{Re}C = \frac{\gamma g R T_0}{(2U_b)^2} \, ,
\end{equation}
where $T_0$ will be specified according to the boundary condition on $\Theta$.
To determine $\beta(t)$, we 
take the spatial average of the $z$-component of (\ref{NSE-perturbation}).
By Gauss's theorem and the divergence-free condition, many terms drop.  Noting also that $\langle u_0\rangle=\langle \Theta_0\rangle =1/2$, the $\beta(t)$ that fixes $\langle u_z\rangle = 0$ is given by
\begin{equation}
    \beta(t) = 
    -C\left(\frac{1}{2}+\langle \Theta\rangle\right)
    - \frac{1}{2} \left. \frac{\partial (u_z)_{00}}{\partial r} \right|_{r=1} \, ,
\end{equation}
where $(\cdot)_{00}$ denotes averaging over $\phi$ and $z$.

\subsection{Fixed temperature difference between bulk and boundary}

We accompany the fixed temperature boundary condition with a fixed bulk temperature $T_b$ in (\ref{eq:Tb}). 
Making the choice 
\begin{equation}
T_0=2\,T_b    
\end{equation}
for the temperature scale, inserting in
(\ref{eq:Cscales}) and rearranging, we find
\begin{equation}
    C_{\Delta T} = \frac{Gr_{\Delta T}}{16\,Re},
    \qquad
    Gr_{\Delta T} = \frac{\gamma\, g\,(T|_{r=R}-T_b)(2R)^{3}}{\nu^{2}}
\end{equation}
wherein we have used the dimensional $T$ of (\ref{eq:tempexpansion}) and subscripted the parameters to clarify that they are based on a temperature difference.  
$Gr_{\Delta T}$ is the Grashof number.

Using the scale $T_0=2T_b$ to non-dimensionalise (\ref{eq:tempexpansion}) and (\ref{eq:Tb}), the dimensionless fluctuations satisfy 
$T|_{r=1}=1$ and
$\langle T\rangle=1/2$. 
As a simple $\Theta_0$ has been chosen that satisfies these conditions, we have that (\ref{T-perturbation}) is accompanied by the boundary condition
$\Theta|_{r=1}=0$ and the condition
$\langle\Theta\rangle=0$.
The latter condition is equivalent to saying that the energy within the domain is constant, and hence the energy entering the domain through the boundary must match the energy extracted by the sink term at each instant.  This sets a value for $a(t)$.   
Taking the spatial average of  (\ref{T-perturbation}) gives         
\begin{equation}
         \label{at-equation-dT}     
        	a(t)=  \frac{4}{RePr} 
            \left( 2 +
            \left.\frac{\partial (\Theta)_{00}}{\partial r}\right|_{r=1} \right) \, .
         \end{equation}
This model has been applied in the simulations of Chu $\textit{et al.}$ \cite{chu2024minimal}.

\subsection{Fixed heat flux at the boundary}

As we already have that $(\partial_r\Theta_0)|_{r=1}=2$
in the decomposition (\ref{eq:Tdecomp2}), we suppose that this is the value of the temperature gradient everywhere, and accompany
(\ref{T-perturbation}) with the boundary condition $(\partial_r\Theta)|_{r=1}=0$.  Using $T_0$ as the temperature scale, the dimensional flux at the wall is everywhere
\begin{equation}
    q_w = 2\,\lambda \frac{T_0}{R}, 
    \qquad \mbox{i.e.}~~
    T_0=\frac{q_w\,R}{2\,\lambda}\,.     
\end{equation}
Inserting this $T_0$ in
(\ref{eq:Cscales}) and rearranging, we find
\begin{equation}
    C_q = \frac{Gr_q}{128\,Re},
    \qquad
    Gr_q = \frac{\gamma \,g\, (2R)^{4}q_w}{\lambda\, \nu^{2}} \, ,
\end{equation}
     where the subscripts have been added to the parameters to indicate that they are based on the heat flux.

The fluctuations may be split into a spatial mean and varying component, $\Theta(\vec{x},t)=(\Theta)_{00}(r,t)+\Theta'(\vec{x},t)$, where $(\cdot)_{00}$ denotes averaging over $\phi$ and $z$.  To the varying components, we apply the boundary condition $(\partial_r\Theta')|_{r=1}=0$. 
The mean component evolves according to the  spatial average of 
(\ref{T-perturbation}), which may be written
\begin{equation}
\label{eq:T-pert-00}
   \frac{\partial (\Theta)_{00}}{\partial t} - \frac{1}{Re\,Pr}
   \nabla^2
   (\Theta)_{00}
   = (N)_{00} - (u_0+(u_z)_{00})\,a(t) .
\end{equation}
We wish the mean component to be consistent with there being a constant background reference temperature in (\ref{eq:Tw}), and therefore apply the boundary condition $(\Theta)_{00}|_{r=1}=0$.  
Note that the temperature can still vary at the boundary, as this condition only fixes the mean value.  
However, it still remains to apply the boundary condition 
$(\partial_r(\Theta)_{00})|_{r=1}=0$,
which is achieved through the variation in $a(t)$.  
Evaluating the radial derivative at the wall gives
         \begin{equation}
         \label{at-equation-q}   
         \left.
        	a(t)= \left.\left( \frac{\partial (N)_{00} }{\partial r} + \frac{1}{RePr}\frac{\partial \nabla^2 (\Theta)_{00} }{\partial r}\right)\right|_{r=1}
        ~\middle/~
            \left. \left( \REF{-2 +} \frac{\partial (u_z)_{00}}{\partial r}
            \right)
            \right|_{r=1} .
        \right.
         \end{equation}

It is worth mentioning that accompanying (\ref{T-perturbation}) with the condition
$(\partial_r\Theta)|_{r=1}=0$ alone, the problem is ill-posed; \REF{see e.g.\ \cite{piller2005direct,baranovskii2024mathematical}}.  The condition $(\Theta)_{00}|_{r=1}=0$ removes non-uniqueness, but note that it cannot be trivially satisfied by evaluating (\ref{eq:T-pert-00}) at the wall ---  $a(t)$ remains undetermined as its coefficient is zero at the wall.

 \subsection{Time-integration code}
    The calculations are carried out by the open-source code openpipeflow.org \citep{willis2017openpipeflow}. 
    Variables are discretised on the domain $\left\{ r,\phi,z \right\}=[0,1]\times[0,2\pi]\times[0,2\pi/\alpha]$, where $\alpha=2\pi/L$, using Fourier decomposition in the azimuthal and streamwise directions and finite difference in the radial direction,
    with points clustered towards the wall.
    An arbitrary variable $f(\vec{x})$ 
    is expanded in the form
    \begin{equation} \label{eq:discr}
      f(r_s,\phi,z)=\sum_{k<|K|} \sum_{m<|M|}
      (f)_{km}(r_s)\,
      e^{i(\alpha kz+m\phi)} \, ,\qquad
      s=1,2,\dots,S\, ,
        \end{equation}
    and the mode $(f)_{00}$ corresponds to the $\phi$- and $z$-average.
    Temporal discretisation is via a second-order predictor-corrector scheme, with an Euler predictor and a Crank-Nicolson corrector applied to the nonlinear terms.  
    The laminar solution is quickly calculated by eliminating 
    azimuthal and axial variations using a 
    resolution $S=64, M=1, K=1$. 
    For a periodic pipe of length
    $L=5D$, the resolution is $S=64, M=76, K=80$ at $Re=5300$, and the resolution is $S=64, M=40, K=44$ at $Re=3000$. For a periodic pipe of length $L=25D$, the resolution is $S=64, M=76, K=400$ at $Re=5300$, and the resolution is $S=64, M=40, K=220$ at $Re=3000$. A time step $\Delta t=0.01$ is adopted. \REF{These resolutions ensure a drop-off of three to four orders of magnitude in the amplitude of the spectral coefficients,
    which experience has shown to be sufficient for accurately simulating shear-turbulence, matching statistics from e.g.\ \cite{eggels1994fully}.  Within the parameter range considered here, the convective state is less computationally demanding to simulate.
}

\section{Results}

In this section, we compare the two different boundary conditions keeping $L=5D$, then we consider the fixed temperature difference boundary condition and compare the flow in $L=5D$ and $L=25D$. Finally we calculate the heat transfer and lifetimes for localised turbulence in the presence of the buoyancy force.
  \subsection{Laminar flow, shear turbulence and convective turbulence}
\begin{figure}
   \includegraphics[width=13.5cm]{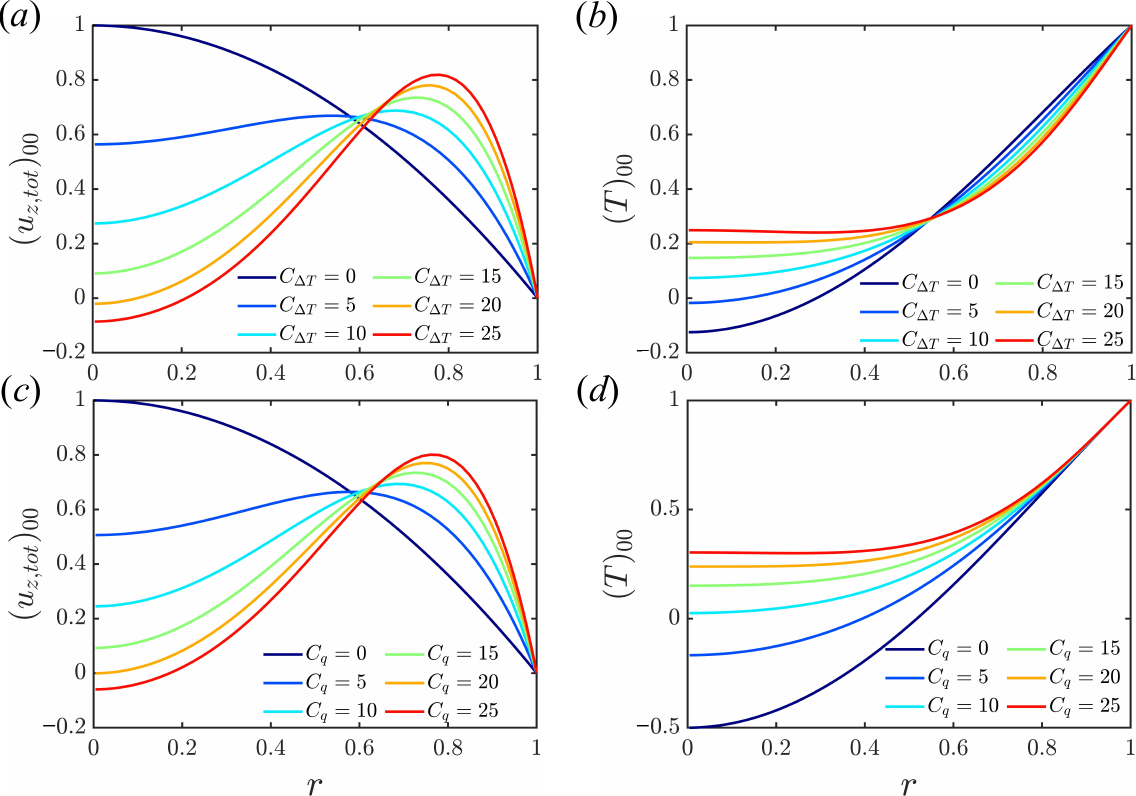}
    \caption{Laminar solution for 
    (\textbf{a-b}) fixed temperature difference; (\textbf{c-d}) fixed boundary heat flux.
    \label{fig:laminar}}
 \end{figure}

  We first verify that the model produces the well-known properties of the laminar solution for both models and for increasing buoyancy parameter $C$ \citep{su2000linear,yoo2013turbulent},
  shown in figure \ref{fig:laminar}.
  Results in figure \ref{fig:laminar} are calculated at $Re=5300$, but laminar profiles are dependent on $C$ and independent of $Re$  \citep{su2000linear}. 
  The laminar velocity profile becomes flattened and even 'M' shaped with the enhancement of heating. Negative velocity near the centre of the pipe at $C_{\Delta T},C_q =20,25$ indicates the occurrence of reversed flow. The laminar temperature profile becomes flattened as $C$ increases. For the fixed temperature difference, an increased temperature gradient near the wall implies increased heat flux and increased Nusselt number $\Nu$, defined in (\ref{eq:Nu}).  For the fixed heat flux case, a reduced temperature difference between the wall and bulk results in increased $\Nu$.

\begin{figure}
   \includegraphics[width=13.5cm]{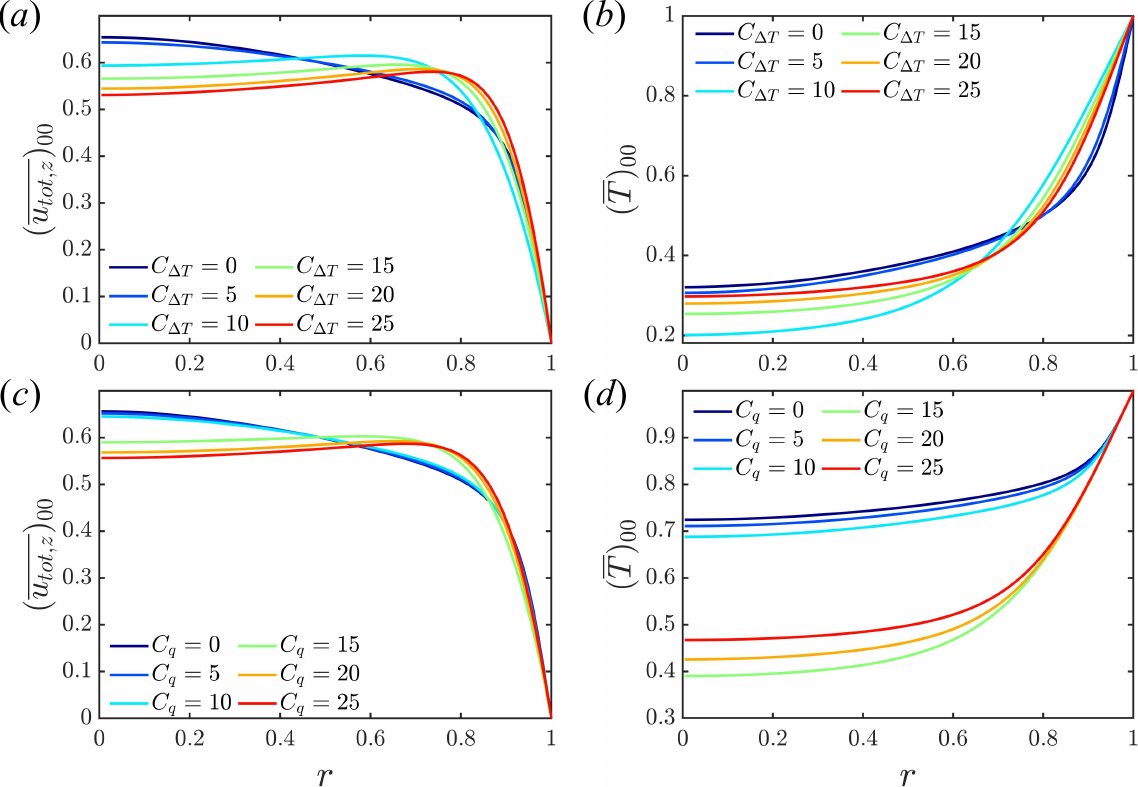}
    \caption{Turbulent mean velocity profiles $\overline{u_{tot,z}}$ and 
    temperature profiles $\overline{T}$
    at $Re=5300, L=5D$: (\textbf{a-b}) 
    fixed temperature difference; (\textbf{c-d}) 
    fixed boundary heat flux. 
    \label{fig:mean}}
 \end{figure}

\begin{figure}
    \includegraphics[width=8cm]{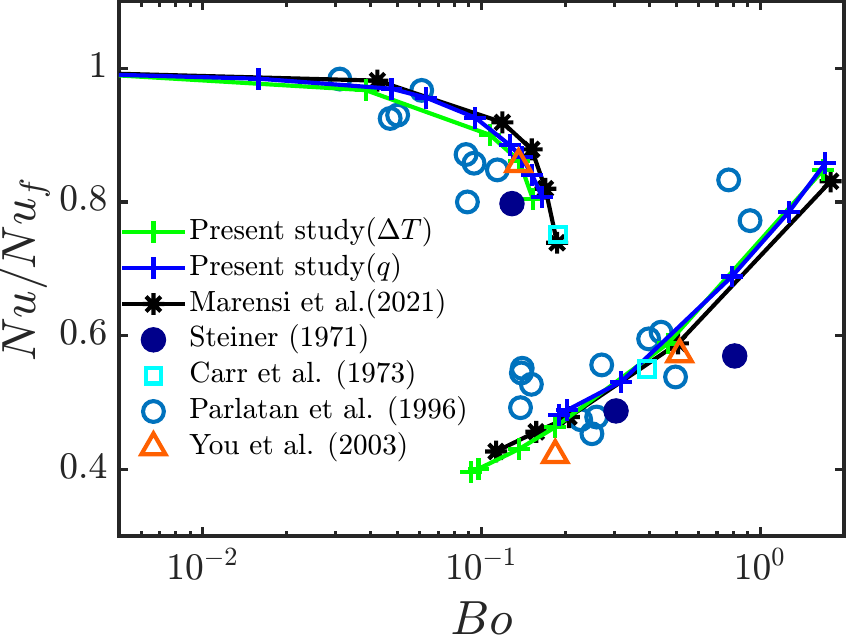}
    \caption{Change in heat flux, normalised by that for the isothermal state ($C \to 0$), as a function of $Bo = 8 \times 10^4 (8\,\Nu \, Gr_{\Delta T})/(Re^{3.425} Pr^{0.8})$ (fixed temperature difference) or $Bo = 8 \times 10^4 (8 Gr_{q})/(Re^{3.425} Pr^{0.8})$ (fixed boundary heat flux). Present data from simulations at $Re = 5300$, $Pr = 0.7$. The upper and lower branches correspond to shear and convective turbulence, respectively.
    Data from \cite{marensi2021suppression,steiner1971reverse,carr1973velocity,
 parlatan1996buoyancy,you2003direct}.
    \label{fig:exp}}
 \end{figure}

 Turbulent mean profiles at $Re=5300$ are shown in figure \ref{fig:mean}. Two regimes are observed in both velocity and temperature profiles, corresponding to shear-driven turbulence and convective turbulence. For the velocity profile, the former state has a flattened shape, while the latter has an 'M' shape due to stronger influence of the buoyancy force.  For these values of $C$, shear turbulence has much greater heat transfer than convective turbulence.
 As $C$ increases, it is observed that heat transfer first becomes weaker, then collapses, and finally, it gradually recovers. 
 This trend is consistent with results reported in the literature \citep{steiner1971reverse,carr1973velocity,parlatan1996buoyancy,you2003direct,zhang2020review}. Both models capture a similar change in heat transfer, but with different critical values of the $C$ parameters. 
Numerical results for the present model are compared with previous numerical results \citep{you2003direct,marensi2021suppression} and experimental results \citep{steiner1971reverse,carr1973velocity,parlatan1996buoyancy}, shown in figure \ref{fig:exp}. 
(Fixed temperature difference and uniform heat sink were adopted in Marensi $\textit{et al.}$ \cite{marensi2021suppression}, while fixed heat flux was applied in You $\textit{et al.}$ \cite{you2003direct}.)
\REF{Averages over at least 4000 time units are used in the calculation of $\Nu$.}
 Two regimes are clearly identified, i.e.\ the heat-transfer deterioration regime and the recovery regime, corresponding to shear turbulence and convective turbulence, respectively.  Both temperature boundary conditions achieve good agreement with experimental results and previous numerical results. 

\begin{figure}
    \includegraphics[width=13.5cm]{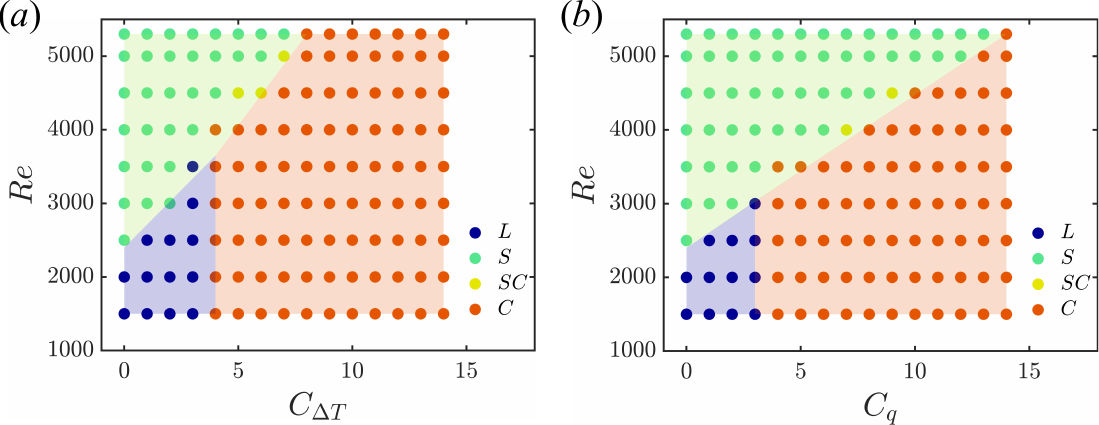}
    \caption{Approximate regions of laminar flow (L), shear turbulence (S) and convective turbulence (C);  SC indicates that the flow may be in either of the two states. (\textbf{a}) Fixed temperature difference; (\textbf{b}) fixed boundary heat flux. 
    \label{fig:SCL}}
 \end{figure}

At lower Reynolds numbers there is a laminarisation regime, seen in figure \ref{fig:SCL}, which shows the approximate regions of the flow states for the two temperature boundary conditions. 
Although there is a difference between the values of $C_{\Delta T}$ and $C_q$ at which transition between different flow regimes occurs, they are consistent in figure \ref{fig:exp}.

\begin{figure}
    \includegraphics[width=13.5cm]{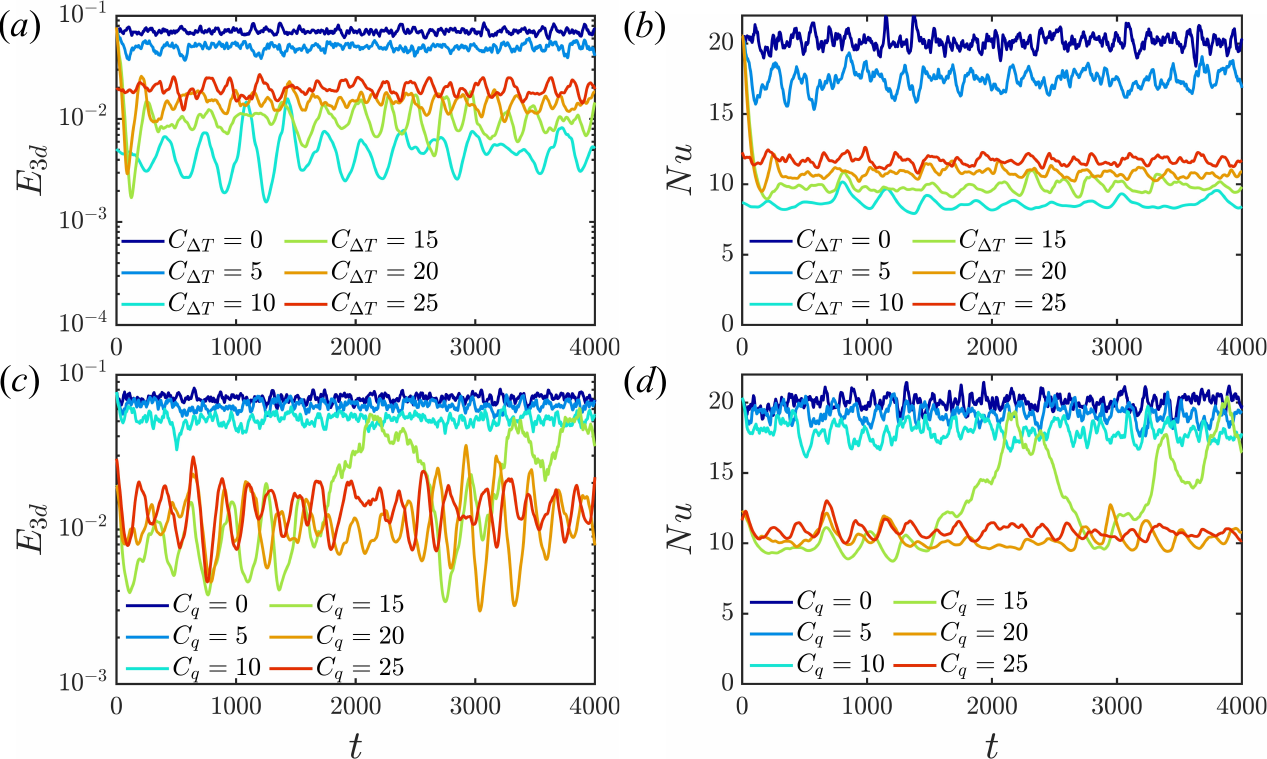}
    \caption{ (\textbf{a,c}) Time series  of $E_{3d}$ (energy of the streamwise-dependent component of the flow) and (\textbf{b,d})  $\Nu(t)$ at different $C_{\Delta T}$ and $C_q$ for $Re=5300$.
    \label{fig:E3d}}
 \end{figure}

  The time evolution of $E_{3d}$ (energy of streamwise-dependent component of the flow) and instantaneous $\Nu (t)$ at different $C_{\Delta T}$ and $C_q$ are presented in figure \ref{fig:E3d}.  Generally, as $C$ is increased, $E_{3d}$ first decreases gradually, then reduces to a much lower energy level at a critical value of $C$, indicating a flow state transition from shear turbulence to convective turbulence \citep{chu2024minimal}. 
  In the convective turbulence state, $E_{3d}$ fluctuates with a much lower frequency.  A clear gap between the shear turbulence regime 
  and the convective turbulence regime (smaller $E_{3d}$ and $\Nu$)  is observed. 
    The critical $C$ is not precise, since close to the border, both states can be observed.
At $Re=5300$, the critical values are 
  $C_{\Delta T} \approx 7$ and 
  $C_q \approx 15$. 
Interestingly, bistability is observed at $C_q=15$, which switches between shear and convective turbulence. In particular, 
the convective state is capable of intermittently triggering bursts of shear-like turbulence,
whereas at lower $Re$ and in isothermal flow, it cannot switch back from the linearly stable laminar state.  

\begin{figure}
    \includegraphics[width=13.5cm]{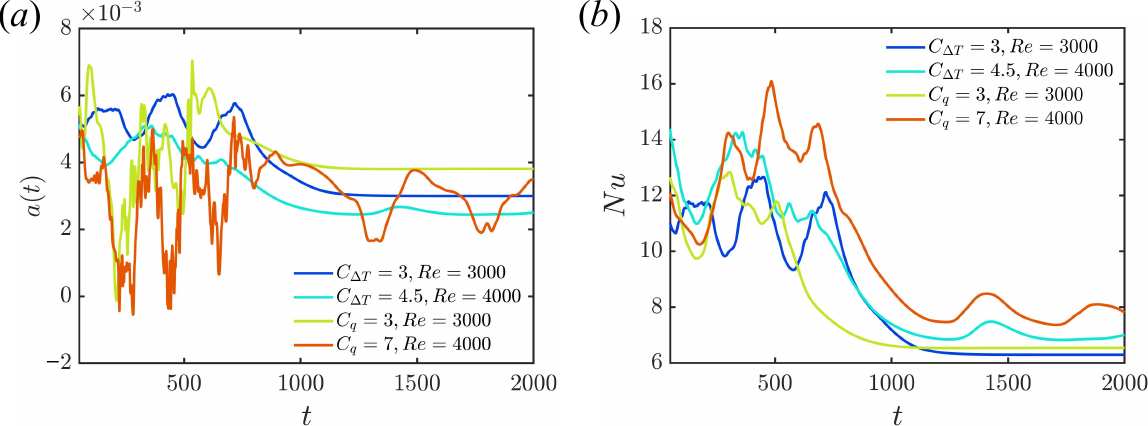}
    \caption{Time evolution of (\textbf{a}) $a(t)$  and (\textbf{b}) instantaneous Nusselt number when shear turbulence collapses to the laminar or convective state for the two boundary conditions.
    \label{fig:at}}
 \end{figure}

 The time evolution of the background temperature gradient $a(t)$ and of $\Nu(t)$ are presented in figure \ref{fig:at} during the transition from shear turbulence to the laminar state, and during the transition from shear turbulence to convective turbulence for the two types of boundary conditions. Transition from shear turbulence to either the laminar state or convective turbulence leads to a reduced Nusselt number.  
 This is accompanied by reduction in the gradient $a(t)$ for the fixed temperature difference model.
 As the heat transfer associated with the new flow is lower, the fluid is heated less and the gradient reduces.
For the fixed heat flux model, however, once the total temperature has adjusted (giving the change in $\Nu$), the time average of $a(t)$ is forced to remain the same so that the heat flux out matches the fixed input flux.
As the energy of the bulk is fixed for the fixed temperature difference, the input and output energies respond immediately to each other, so that $a(t)$ and $\Nu(t)$ vary together.
For the fixed flux condition, $\Nu(t)$ varies due to differences in the bulk temperature, which responds in a time-integrated fashion relative to the heat flux out.  Hence fluctuations in $\Nu(t)$ are less rapid than those in $a(t)$ for the fixed flux boundary condition.

\begin{figure}
   \includegraphics[width=13.5cm]{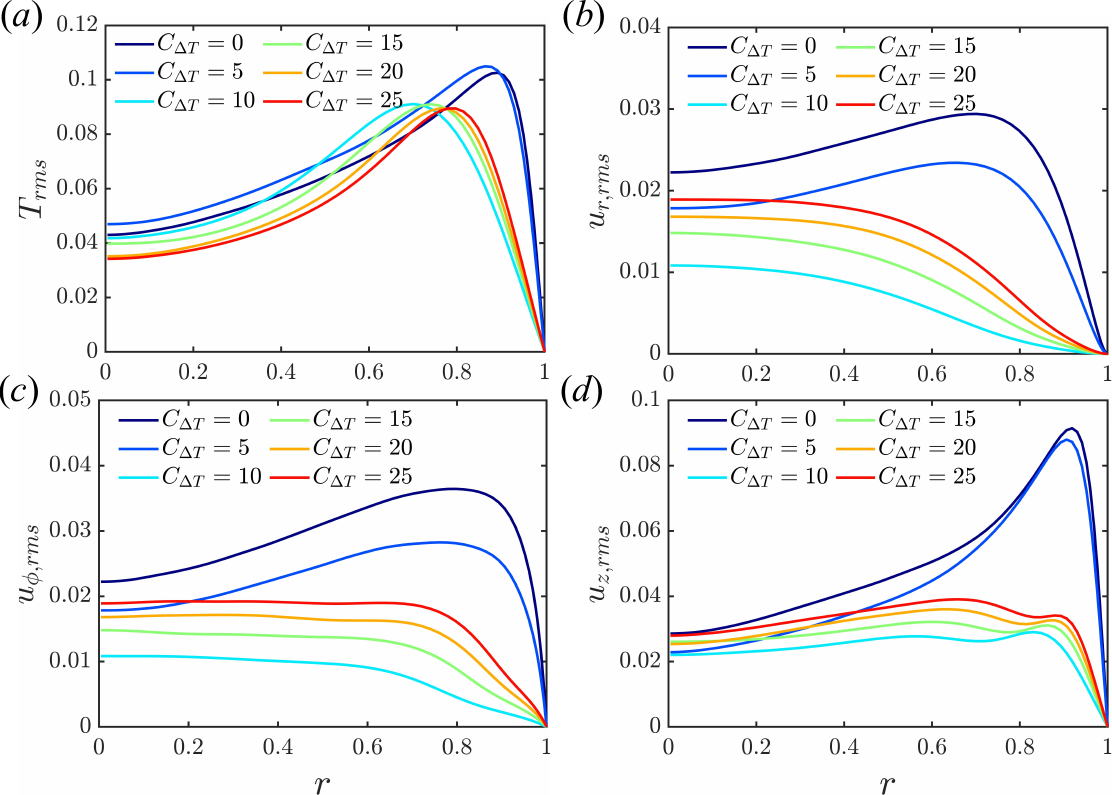}
    \caption{Profiles of RMS temperature and velocity fluctuations at $Re=5300, L=5D$: (\textbf{a}) ${T_{rms}}$; (\textbf{b}) ${u_{r,rms}}$; (\textbf{c}) ${u_{\phi,rms}}$; 
    (\textbf{d}) ${u_{z,rms}}$. Fixed temperature difference.
    \label{fig:rmsdt}}
 \end{figure}

\begin{figure}
   \includegraphics[width=13.5cm]{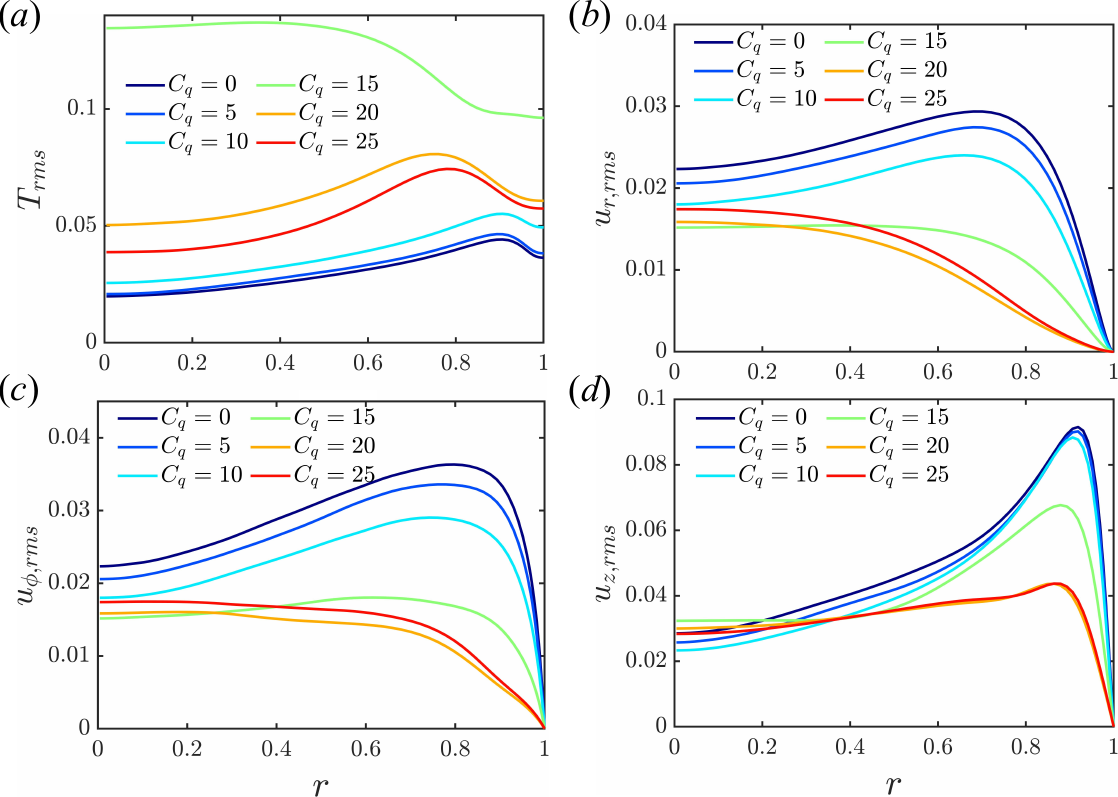}
    \caption{The profile of root mean square of temperature and velocity at $Re=5300, L=5D$: (\textbf{a}) ${T_{rms}}$; (\textbf{b}) ${u_{r,rms}}$; (\textbf{c}) ${u_{\phi,rms}}$; (\textbf{d}) ${u_{z,rms}}$. Fixed boundary heat flux.
    \label{fig:rmsq}}
 \end{figure}

Root-mean-square (RMS) deviations from $\overline{(u_{z,tot})_{00}}$ and $\overline{(T)_{00}}$ are shown in
figure \ref{fig:rmsdt} and figure \ref{fig:rmsq} for the fixed temperature and fixed flux boundary conditions respectively, using data from $t=1000$ to $t=4000$ for each simulation.
Interestingly, there are two peaks of streamwise velocity fluctuation observed in convective turbulence when the fixed temperature difference is adopted, see figure \ref{fig:rmsdt}($\textit{d}$). At $C_{\Delta T}=10$, the peak near the wall dominates, while the peak far away from the wall is larger at $C_{\Delta T}=25$. The two peaks are in good agreement with You $\textit{et al.}$ \cite{you2003direct} in figure 4 and Cruz $\textit{et al.}$ \cite{cruz2024high} in figure 3. 
The main difference between the two models is in the temperature fluctuations $T_{rms}$.  As only the mean temperature at the wall is fixed for the fixed flux model,  fluctuations are possible even at the wall.  $T_{rms}$ for $C_q=15$ is especially large, due to the bistability mentioned earlier (see figure \ref{fig:E3d}($\textit{c,d}$)).
Otherwise, the results are similar, and 
differences between the shear and convective regimes are observed in the RMS fluctuations for both models.
In the shear turbulence regime, 
the peak of temperature fluctuation is close to the wall and moves away from the wall with increased heating.  In the convective turbulence regime, the peak of the temperature fluctuations is much further away from the wall, and moves closer to the wall again as the heating is increased. 
The peak fluctuations for all velocity components are close to the wall in the shear turbulence regime, and weaken as $C$ increases.  For the convective regime, fluctuations are spread more evenly across the domain and strengthen as $C$ increases further. Results are consistent with other calculations of RMS quantities \citep{you2003direct,zhao2018direct,cruz2024high}.


\subsection{Short vs long periodic pipes}

\begin{figure}
    \includegraphics[width=13.5cm]{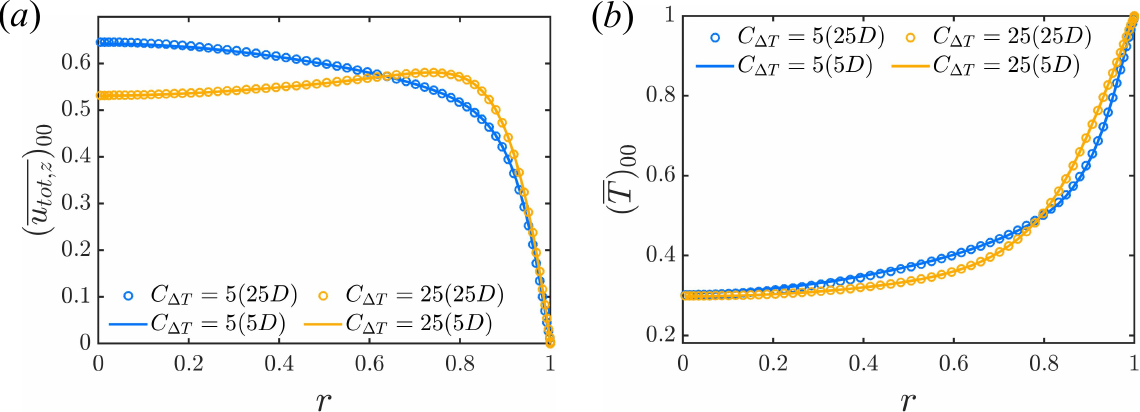}
    \caption{Comparison of mean (\textbf{a}) streamwise velocity and (\textbf{b}) temperature profile between short periodic pipe ($L=5D$) and long periodic pipe ($L=25D$). Two typical flow states are simulated, i.e. shear turbulence ($C_{\Delta T}=5$) and convective turbulence ($C_{\Delta T}=25$) at $Re=5300$.
    Fixed temperature difference boundary condition is used. 
    \label{fig:longpipe}}
 \end{figure}

\begin{figure}
    \includegraphics[width=13.5cm]{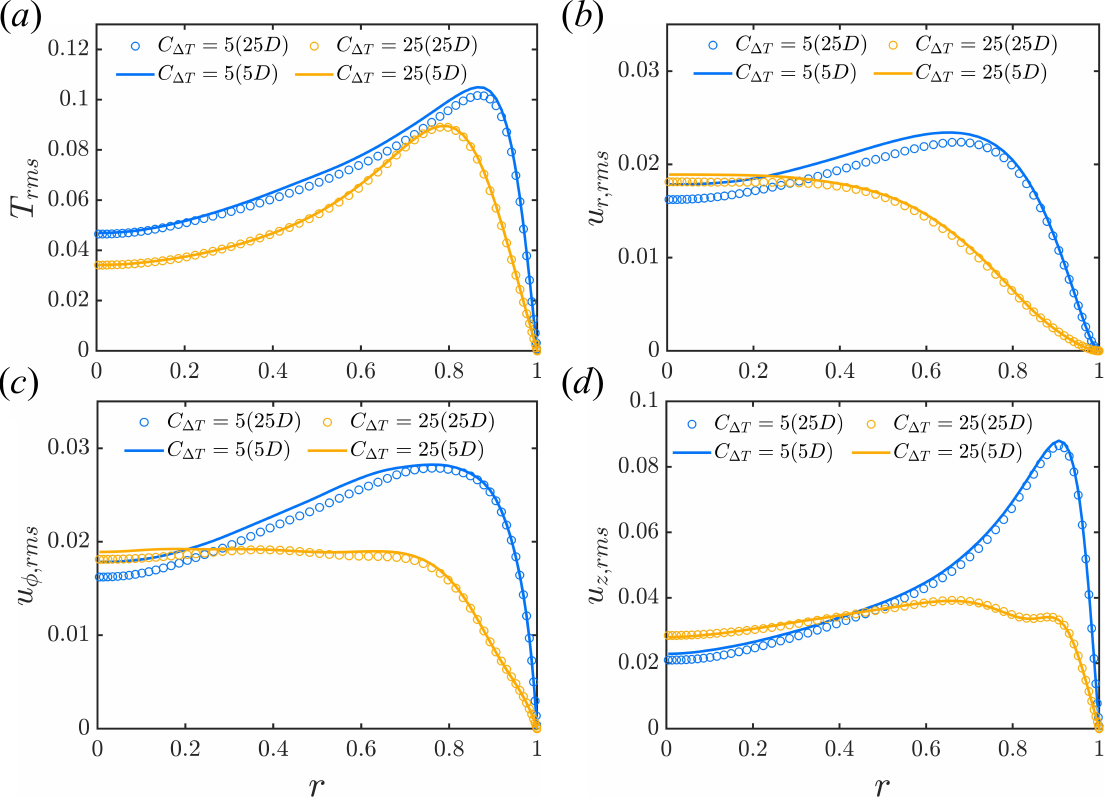}
    \caption{Comparison of (\textbf{a}) ${T_{rms}}$, (\textbf{b}) ${u_{r,rms}}$,(\textbf{c}) ${u_{\phi,rms}}$ and (\textbf{d}) ${u_{z,rms}}$ between short periodic pipe ($L=5D$) and long periodic pipe ($L=25D$). Two typical flow states are simulated, i.e. shear turbulence ($C_{\Delta T}=5$) and convective turbulence ($C_{\Delta T}=25$) at $Re=5300$.  Fixed temperature difference. 
    \label{fig:longpiperms}}
 \end{figure}

As the two models give consistent results, only the fixed temperature difference model will be considered here.
The axially periodic boundary condition 
 could impose some difference in results compared to true flow.  
 Thus, here we use a longer pipe, $L=25D$, for comparison.
 Figure \ref{fig:longpipe} shows the mean velocity and temperature profiles of a short pipe ($L=5D$) and a longer pipe ($L=25D$) in shear turbulence regime ($C_{\Delta T}=5$) and strong convective turbulence ($C_{\Delta T}=25$). The results for the two pipe lengths are in good agreement, suggesting that $L=5D$ is enough for capturing the mean profiles. The distributions of RMS of temperature and velocity for the short pipe and long pipes are shown in figure \ref{fig:longpiperms}.  
 There are some small differences, but the agreement is still good.  Differences are smaller for convective turbulence.  For shear turbulence, there is a little deviation in the centre of the pipe for the cross-stream velocity components.
The results in the near-wall region are well matched, suggesting that simulations in a short pipe are expected to capture the heat transfer processes accurately.  
 
\begin{figure}
    \includegraphics[width=13.5cm]{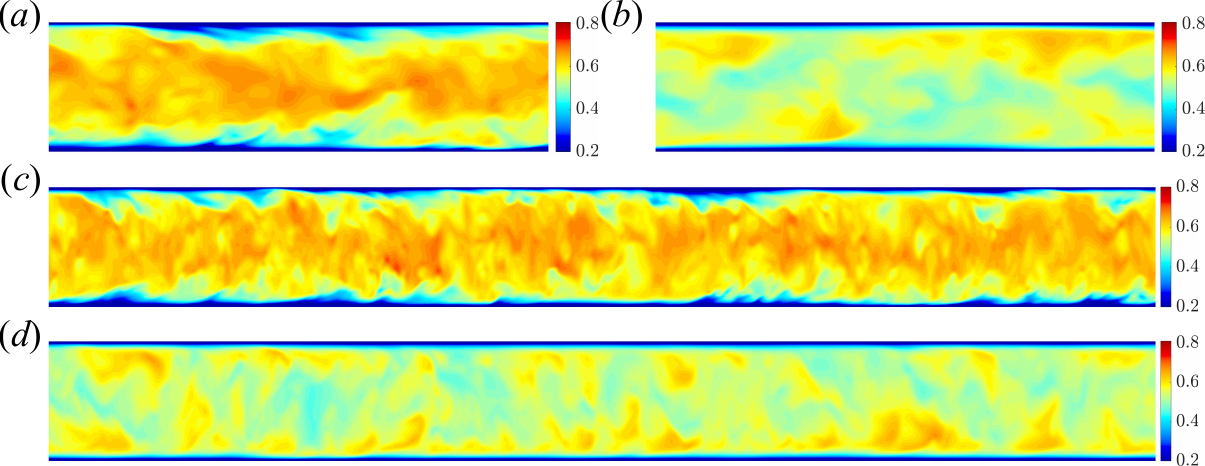}
    \caption{ Contours of streamwise velocity in $rz$ cross-section for shear turbulence ($C_{\Delta T}=5$) in ($\textbf{a}$) $L=5D$ ($\textbf{c}$) $L=25D$, and convective turbulence ($C_{\Delta T}=25$) in ($\textbf{b}$) $L=5D$ ($\textbf{d}$) $L=25D$ at $Re=5300$. 
    For the long pipe, the $z$-axis has been scaled to show the whole pipe.
    \label{fig:contour}}
 \end{figure}

\begin{figure}
    \includegraphics[width=13.5cm]{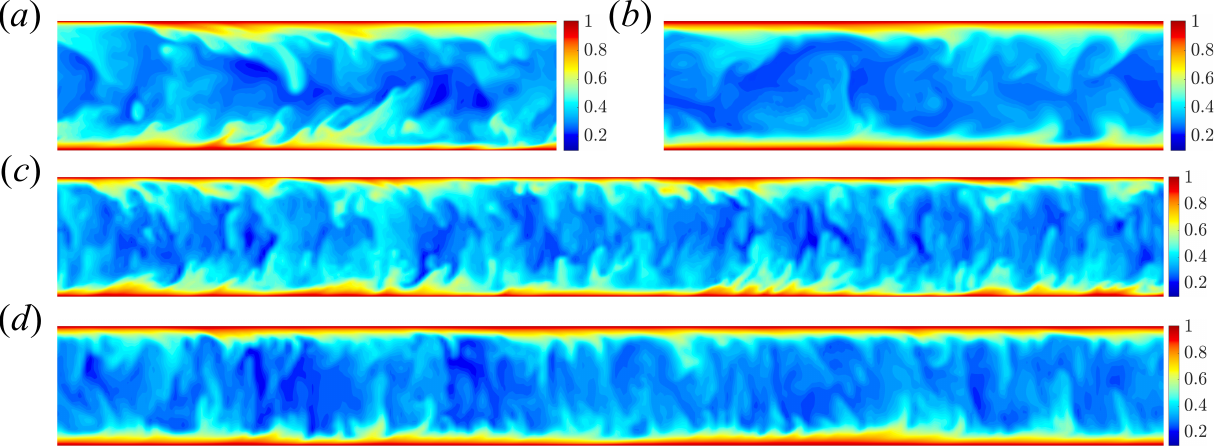}
    \caption{ Contours of temperature in $rz$ cross-section for shear turbulence ($C_{\Delta T}=5$) in ($\textbf{a}$) $L=5D$ ($\textbf{c}$) $L=25D$, and convective turbulence ($C_{\Delta T}=25$) in ($\textbf{b}$) $L=5D$ ($\textbf{d}$) $L=25D$ at $Re=5300$. 
        For the long pipe, the $z$-axis has been scaled to show the whole pipe.
    \label{fig:Tcontour}}
 \end{figure}

 Contours of streamwise velocity and temperature in the $rz$-cross-section for the two pipe lengths are shown in figure \ref{fig:contour} and figure \ref{fig:Tcontour} respectively. The difference in velocity between shear turbulence (figure \ref{fig:contour}($\textit{a,c}$)) and convective turbulence (figure \ref{fig:contour}($\textit{b,d}$)) is clear: shear turbulence 
 has strong low-speed regions near the wall (associated with streaks).  These are essentially absent in convective turbulence, and are replaced with localised regions of fast flow near the wall, while the core flow moves more slowly.
No obvious difference in the contour plots is observed between short and long pipes, for both velocity and temperature fields. 
 
\begin{figure}
    \includegraphics[width=13.5cm]{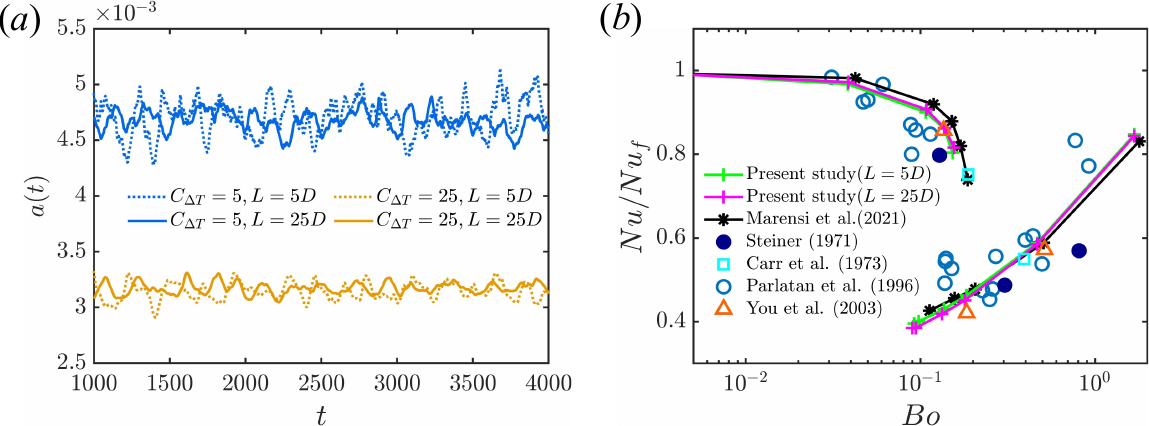}
    \caption{(\textbf{a}) Time evolution of $a(t)$ for short ($L=5D$)  and longer pipe ($L=25D$). (\textbf{b}) Normalised  Nusselt number for the short and longer pipe at $Re=5300$.
    $Bo$ defined as in figure \ref{fig:exp}.  Fixed temperature difference. 
    \label{fig:exp1}}
 \end{figure}

 Time evolution of $a(t)$ for the two pipe lengths is shown in figure \ref{fig:exp1}(\textit{a}).  Curves at matching $C_{\Delta T}$ are quite close, but smaller fluctuations in $a(t)$ are observed for the longer pipe. This is expected, as the larger domain gives more steady volume-averaged quantities used in the calculation of $a(t)$.  Nusselt numbers for the short and long pipes at several $C_{\Delta T}$ are compared in figure \ref{fig:exp1}(\textit{b}). There is almost no difference in Nusselt number over a wide range $C_{\Delta T}$ covering both shear turbulence and convective turbulence. Therefore, it is concluded that the simulation of a short periodic pipe ($L=5D$) is enough to predict heat transfer and flow dynamics for fully turbulent flow. 
 For $C$ close to critical, however, data from either one state or the other was used in the calculation of $\Nu$, so that intermittency is not fully accounted for.  We consider this next.


In isothermal flow, localised turbulent
patches are called puffs and slugs.
\citep{avila2023transition}. 
Puffs appear for $Re \approx 1800$ and are
statistically steady in axial extent.
From $Re\gtrsim 2250$, they start to expand and are called slugs.  
However, within the expanding turbulent region (that will eventually fill a periodic domain), laminar patches remain present for $Re$ up to approximately $2800$ \citep{marensi2021onset}.
Thus, there is a large range over which the intermittent nature of turbulence 
cannot be captured in a short periodic domain of length $L=5D$.
Puffs and slugs have frictional drag between values extrapolated from the fully turbulent or laminar regimes, and are marked as a hatched area in the Moody
diagram \citep{moody1944friction}.  
For a heated pipe, this will affect estimations of $\Nu$.

\begin{figure}
    \includegraphics[width=13.5cm]{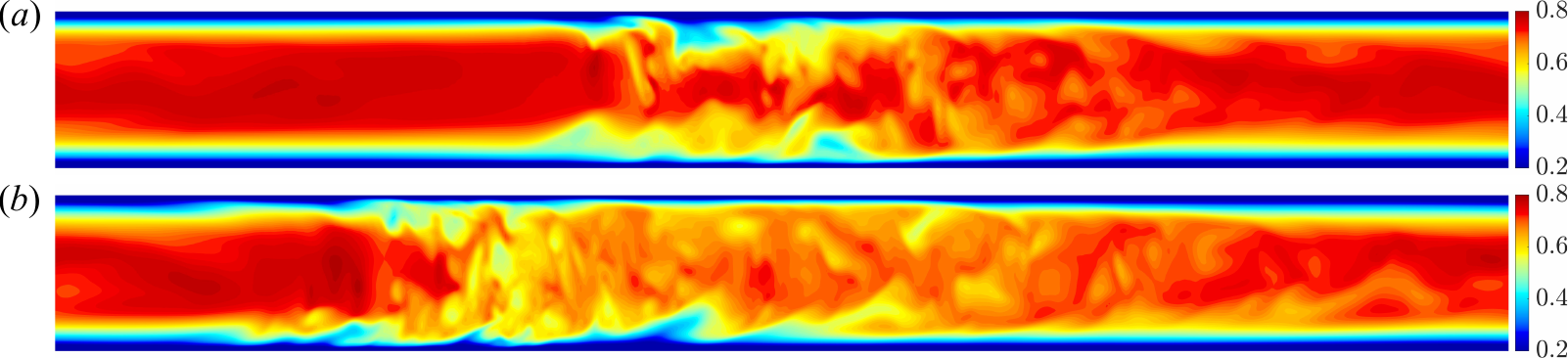}
    \caption{Contour of streamwise velocity in long pipe ($L=25D$ at $Re=3000, C_{\Delta T}=1.9$): $(\textbf a)$ puff and $(\textbf{b})$ slug. 
    \label{fig:puffslug}}
 \end{figure}

\begin{figure}
    \includegraphics[width=7cm]{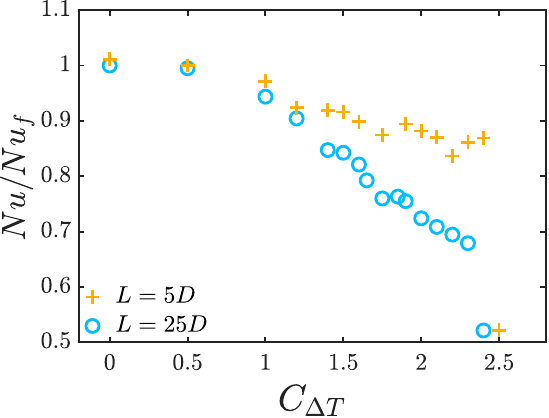}
    \caption{Comparison of Nusselt number 
    for transitional $C_{\Delta T}$ for 
    a short and long periodic domain ($L=5D,\,25D$) at $Re=3000$. 
   Values for (intermittent) turbulence are shown, except for the final two laminar points.
    \label{fig:Nupuff}}
 \end{figure}

In vertical heated pipe flow, intermittent turbulence exists around the boundary between laminar and shear turbulence at higher Reynolds numbers, at the meeting of green and blue regions in figure \ref{fig:SCL}. Examples of puff and slug at $Re=3000,C_{\Delta T}=1.9$ are shown in figure \ref{fig:puffslug}. 
Nusselt numbers for the short and long pipes at $Re=3000$ are shown in figure \ref{fig:Nupuff}. 
At small $C_{\Delta T}$ there is almost no difference, as the turbulence fills the pipe.  As $C_{\Delta T}$ increases, the difference in $\Nu$ between the short and long pipe becomes substantial, due to the appearance of localised turbulence.  Eventually, laminarisation occurs, marked by the final two points.

\subsection{The lifetime of localised shear turbulence}
 The mean lifetimes of turbulent puffs in isothermal flow, and its scaling with Reynolds number, have been closely investigated
\citep{willis2007critical,willis2009turbulent,hof2008repeller,avila2010transient}. At each $Re$, the mean lifetime must be estimated from a series of simulations 
or experiments, and data is often truncated,
due to limited simulation time or finite length of pipe
\citep{avila2010transient}. 
To examine whether the lifetime of puffs in heated pipes behave similarly, we calculated the survivor functions at $Re=3000$ for several $C_{\Delta T}$.  To generate the initial conditions for the simulations, a localised disturbance was applied to the laminar Poiseuille flow at $Re = 3000, C_{\Delta T}=1.9$ and the resulting puff was evolved for $t \approx 5000 $. Snapshots of the full velocity field were recorded every $20$ time units, generating a large collection of initial conditions.  Subsequently, simulations at larger $C_{\Delta T}$ were performed starting from these initial conditions and were monitored until the flow laminarised. The
 criterion for laminarization was $E_{3d}<10^{-3}$, below which turbulent motions have decayed beyond recovery. 
 
\begin{figure}
    \includegraphics[width=13.5cm]{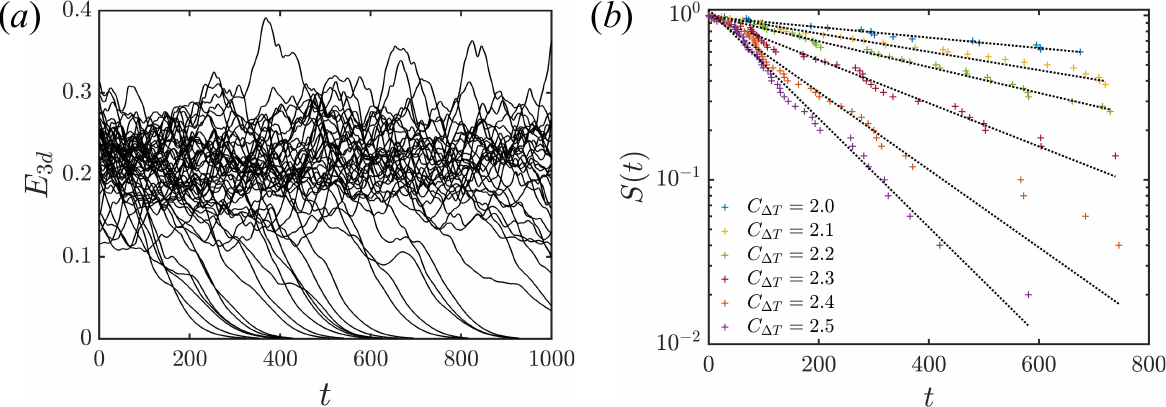}
    \caption{The time evolution of $(\textbf{a})$ $E_{3d}$ of $50$ arbitrary initial turbulent fields at $C_{\Delta T}=2$. $(\textbf{b})$ Survivor function for several values of the buoyancy parameter.   $n = 50$ samples for each case. $L=25D$. 
    \label{fig:lifetime}}
 \end{figure}

 The time evolution of $E_{3d}$ of $n=50$ arbitrary initial turbulent fields at $C_{\Delta T}=2$ are shown in figure \ref{fig:lifetime}(\textit{a}). 
 Some cases decay to the laminar state, while others remain turbulent for the period of the simulation.  The decay of turbulence leads to a large drop in the Nusselt number and an exponential decay of $E_{3d}$, so that laminarisations are clearly identifiable. 
For a finite set of samples, the survivor function is approximated by
 \begin{equation}
         \label{eq:St}   
         S(t)=\frac{r}{n},
         \end{equation}
where $r$ is the number of puffs that survive up to time $t$.  For example, all initial conditions survive before $t=10$, then $S(t)=1 $ when $t<10$. In this way, we can calculate the lifetime of survivor probability from $1$ to $\frac{1}{50}$.
However, due to the finite time it takes for $E_{3d}$ to drop to $10^{-3}$, the data in figure \ref{fig:lifetime}(\textit{b}) have been shifted to the left by the time of the earliest decay ($\approx 250$).
As $C_{\Delta T}$ increases, the mean lifetime of puffs decreases.  The distributions remain exponential in form for each $C_{\Delta T}$.  This 
 indicates that puff decay induced  by heating is also a memoryless process, 
 corresponding to the escape from a strange saddle \citep{willis2007critical,avila2010transient}.  The enhancement of heating has a similar effect to that of the decrease in Reynolds number in isothermal flow. 

\section{Conclusions}

In this work we have presented a 
derivation of a model for vertically heated pipe flow 
that includes a time-dependent axial temperature gradient.  This gradient adjusts in response the the flow pattern.  A transition from shear turbulence to convective turbulence is well-known to lead to a drop in heat transfer.  For the fixed temperature model, reduced heat transferred into the fluid leads to a reduction in the temperature gradient.  With the fixed heat flux boundary condition, however, as the energy withdrawn from the domain is proportional to the gradient, the gradient is forced to remain the same on average to match the energy entering the domain.

Laminar solutions are calculated numerically 
for several values of the buoyancy parameter, and are consistent with previous reports \citep{su2000linear,you2003direct,marensi2021suppression}. 
For turbulent flow,
the time-averaged velocity and temperature profiles, and their RMS fluctuations, are calculated for both boundary conditions. 
The two turbulent regimes, i.e.\ shear turbulence and convective turbulence, are easily distinguished in the mean profiles and RMS flucutations.
The dependence of flow state over the space of $Re$ and $\Nu$ is calculated for both boundary conditions along with various statistics.  Statistics show minor differences between boundary conditions, but both show good consistency with previous calculations and experiments.  Of particular interest are the RMS temperature fluctuations, as they can be non-zero at the wall for the fixed-flux case.  
Also of interest is that convective turbulence can trigger bursts of shear-like turbulence when close to the critical $C$ between the two states.
(For isothermal flow, shear-turbulence cannot return from the linearly stable laminar state.)


Further simulations are carried out to examine the effect of the periodic length of the pipe on the turbulent statistics and heat transfer.  The short pipe $L=5D$ and long pipe $L=25D$ show almost no difference in the mean velocity and mean temperature profiles.  However, there are some minor mismatches in the RMS of velocity and temperature. The length of the pipe was found to have more effect on shear turbulent state,
possibly due to spatial intermittency.
The mismatch mainly appears in the centre of the pipe, while there is always a good agreement in the near-wall regime.  Hence, the short pipe still captures accurate Nusselt numbers, provided that the flow is not too intermittent.  
In that case, simulations with a short pipe are likely to overestimate heat transfer, as shear turbulence fills the domain. 

Finally, we have recorded the lifetime of the localised turbulence with heating, confirming it also follows a memoryless process corresponding to the escape from a strange saddle.  
Using the 
previous model of \cite{marensi2021suppression} close to criticality at larger $Re$, strong fronts and puffs have been found to disappear \cite{zhuang2023discontinuous}.

{The code used for these calculations is available at openpipeflow.org.}



\vspace{6pt}

\funding{S.C. acknowledges funding from Sheffield–China Scholarships Council PhD Scholarship
Programme (CSC no. 202106260029).}


\informedconsent{Informed consent was obtained from all subjects involved in the study.}

\conflictsofinterest{The authors declare no conflicts of interest.} 



\abbreviations{Abbreviations}{
The following abbreviations are used in this manuscript:\\

\noindent 
\begin{tabular}{@{}ll}
RMS & Root Mean Square\\
DNS & Direct Numerical Simulation
\end{tabular}
}




\begin{adjustwidth}{-\extralength}{0cm}

\reftitle{References}



\bibliography{ref}

%


  \PublishersNote{}
  \end{adjustwidth}
\end{document}